\documentclass[a4paper,11pt]{article}
\pdfoutput=1 

\usepackage{docstyle} 

\usepackage[T1]{fontenc} 

\usepackage{lmodern}
\usepackage{enumitem}
\usepackage{slashed}

\title{\textcolor{mycolor}{Ghost-Free Electroweak Symmetry Breaking\\with Weakly Nonlocal Interactions}}

\author[a]{Pratik Chattopadhyay}
\author[b]{and Florian Nortier}

\affiliation[a]{Namtech, IIT Gandhinagar campus, Gandhinagar 382055, India}
\affiliation[b]{Université Claude Bernard Lyon 1, CNRS/IN2P3, IP2I Lyon, UMR 5822,\\Villeurbanne, F-69100, France}

\emailAdd{pratikpc@gmail.com}
\emailAdd{f.nortier@ip2i.in2p3.fr}

\abstract{Weakly nonlocal (WNL) Quantum Field Theories (QFT's) may define a new class of UV-completions in particle physics and gravity, without introducing any new elementary particle. One problematic issue is how to realize spontaneous symmetry breaking without introducing an infinite tower of ghosts in the perturbative spectrum. In this article, a WNL extension of the Standard Model (SM) is proposed: the Fuzzy Standard Model (FSM). It is a smooth deformation of the SM based on covariant star-products of fields. This new formalism realizes electroweak symmetry breaking without ghosts at tree-level. We give evidences that the FSM exhibits Vainshtein screening, aka classicalization, in the deep-UV. This could solve the electroweak hierarchy problem if it occurs at the TeV-scale.}


\arxivnumber{2311.08311}

\begin{document}

\maketitle
\flushbottom

\section{Introduction}
\label{introduction}
The electromagnetic and subatomic interactions are well described by the Standard Model (SM) of particle physics \cite{Halzen:1984mc}, which is a local Quantum Field Theory (QFT). However, gravity is argued to have some nonlocal features \cite{Giddings:2011xs}, which may tell us that the fundamental theory of Nature has some weaker notion of locality than textbook QFT, like String Theory \cite{Marshakov:2002ff}.

During the last decade, there has been a revival of the old program\footnote{For a review on the pre-SM attempts to define nonlocal UV-deformations of QFT's, cf. the book \cite{Namsrai:1986md}. See also the pioneer works by Efimov in Refs.~\cite{Efimov:1970jvp, Efimov:1971hgl, Alebastrov:1971sij}.} of building Lorentz-invariant weakly nonlocal (WNL) QFT's\footnote{The nomenclature ``infinite-derivative QFT'' is also widely spread in the modern literature.} \cite{BasiBeneito:2022wux}. When applied to Yang-Mills or gravity theories, one usually uses the Krasnikov-Terning gauge-covariant nonlocal scheme \cite{Krasnikov:1987yj, Terning:1991yt}. Such theories are claimed to have different perturbative or nonperturbative UV-behaviors, depending on the choice of the WNL form factors, like superrenormalizability/UV-finiteness à la Kuz'min-Tomboulis \cite{Kuzmin:1989sp, Tomboulis:1997gg, Modesto:2011kw, Modesto:2014lga, Modesto:2015lna, Modesto:2015foa, Calcagni:2023goc}, UV/IR duality by worldline inversion symmetry \cite{Abel:2019ufz, Abel:2019zou} (to copy stringy modular symmetry \cite{Abel:2021tyt, Abel:2023hkk}), or UV-obstruction by classicalization \cite{Dvali:2010bf, Dvali:2010ue, Dvali:2010jz, Dvali:2010ns, Dvali:2011nj, Dvali:2011th, Grojean:2011bq, Dvali:2012zc, Dvali:2012mx, Dvali:2014ila, Keltner:2015xda, Addazi:2015ppa, Addazi:2018ivg, Dvali:2020wqi, Addazi:2020nkm, Dvali:2022vzz, Buoninfante:2023dyd}. Furthermore, inspired by tower truncation in String Field Theory \cite{Ohmori:2001am}, several groups have investigated effective toy models with exponential form factors \cite{Buoninfante:2018mre}, where one tries to ``mimic'' the effect of the string scale in gravity  \cite{Biswas:2011ar, Biswas:2013cha, Calcagni:2014vxa} and particle physics \cite{Biswas:2014yia, Ghoshal:2017egr, Gama:2018cda, Hashi:2018kag, Buoninfante:2018mre, Buoninfante:2018gce, Ghoshal:2018gpq, Ghoshal:2020lfd, Frasca:2020jbe, Krasnikov:2020kgh, Frasca:2020ojd, Frasca:2021iip, Su:2021qvm, Nortier:2021six, Mo:2022szw, Ghoshal:2022mnj, Buoninfante:2022trw, Krasnikov:2022xsi, Capolupo:2022awe, Chatterjee:2023ehr, Abu-Ajamieh:2023syy, Capolupo:2023kuu, Abu-Ajamieh:2023roj, Nortier:2023dkq, Abu-Ajamieh:2023txh}, but the actual perturbative control can be questioned \cite{Talaganis:2014ida}. There is also an extensive literature on applications in black hole physics and cosmology (cf. reviews \cite{Buoninfante:2022ild, Koshelev:2023elc}). The form factors are usually chosen ghost-free \cite{Buoninfante:2018mre, BasiBeneito:2022wux}, otherwise one gets trouble with stability or unitarity \cite{Woodard:2015zca, Platania:2022gtt, Kubo:2023lpz}, and perturbative analyticity/unitarity has been shown (at least) for UV-finite scalar models, and expected to hold with gauge theories when the UV-behavior is under perturbative control \cite{Efimov:1966ylf, Tomboulis:1997gg, Pius:2016jsl, Carone:2016eyp, Briscese:2018oyx, Chin:2018puw, Pius:2018crk, DeLacroix:2018arq, Briscese:2021mob, Koshelev:2021orf, Buoninfante:2022krn}.

However, a well-known difficulty of this program arises when a scalar field acquires a vacuum expectation value (vev): an infinite tower of ghosts pops up above the WNL scale in the physical vacuum \cite{Barnaby:2007ve, Galli:2010qx, Gama:2018cda, Hashi:2018kag, Koshelev:2020fok, Nortier:2023dkq}. This is a serious issue since, in the Glashow-Weinberg-Salam (GWS) model of electroweak (EW) interactions \cite{Glashow:1961tr, Weinberg:1967tq, Salam:1968rm}, spontaneous electroweak symmetry breaking (EWSB) occurs via tachyon condensation, aka the Higgs mechanism \cite{Englert:1964et, Higgs:1964pj, Guralnik:1964eu}. Fortunately, 3 different possibilities have been proposed to realize a WNL Higgs mechanism without ghosts: gauge-Higgs unification \cite{Hashi:2018kag}, tree-duality \cite{Modesto:2021okr, Modesto:2021soh} and a covariant star-product formalism \cite{Nortier:2023dkq}.

In this article, we continue to investigate the framework of covariant star-products between fields, which has been recently proposed by one of the authors in Ref.~\cite{Nortier:2023dkq} for an Abelian Higgs model without fermions. Our goal is to generalize the formalism to include non-Abelian gauge theories, fermions and Yukawa couplings, in order to build a minimal version of the Fuzzy Standard Model (FSM): a WNL deformation of the SM that is ghost-free in the physical EW vacuum. We stay mostly agnostic about the precise choice of the WNL form factors involved in the star-products: it is of course important to study the UV-behavior with quantum corrections, but here we restrict ourselves to a tree-level analysis. N.B.: Our conventions are the same as in Peskin \& Schroeder's textbook \cite{Peskin:1995ev}.

\section{Bosonic Fields}
\label{bosonic_fields}
\subsection{Pure Gauge Sector}
One wants to define covariant star-products\footnote{We use $\bullet$ subscripts to introduce in a generic way symbols that can have different subscripts.} $\star_\bullet$ between fields, as well as their noncovariant avatars $\ast_\bullet$. One introduces entire functions $\vartheta_\bullet(z)$ and fuzzy-plaquettes $\eta_\bullet = \left( 1/\Lambda_\bullet \right)^2$, with the WNL scales $\Lambda_\bullet$. They are used to define ghost-free form factors (entire functions without zeroes in $\mathbb{C}$) that can be in general written and expanded as
\begin{equation}
e^{\vartheta_\bullet(z)} = \sum_{n=0}^{+\infty} c_{\bullet}^{(n)} z^{n} \, , \ \ \ 
c_{\bullet}^{(n)} \in \mathbb{R} \, , \ \ \ 
\vartheta_\bullet(0) = 0 \, .
\label{form_factor_exp}
\end{equation}
Such a form factor is WNL, i.e. in the local limit $\eta_\bullet \rightarrow 0$, one retrieves Dirac distributions $\delta(z)$. With the Minkowski metric $g_{\mu\nu}$, one introduces the tensors $\eta_\bullet^{\mu \nu} = \eta_\bullet g^{\mu \nu}$ in the definitions of the star-products.

The weak hypercharge bosons $B_\mu$ -- in the gauge representation\footnote{We use the notation $\left( R_C, R_W \right)_Y$ for a representation of the SM gauge group $SU(3)_C \times SU(2)_W \times U(1)_Y$, where the subscripts $C$, $W$ and $Y$ refer to color, weak isospin and weak hypercharge, respectively.} $\left( \mathbf{1}, \mathbf{1} \right)_0$ -- are the gauge bosons of the group $U(1)_Y$ of coupling $g_1$. The gauge field strength tensor is $\mathcal{B}_{\mu\nu} = \partial_\mu B_\nu - \partial_\nu B_\mu$. One defines the star-products\footnote{This star-product is nonassociative in general, and one needs to keep this property in mind when dealing with operators involving more than 1 star-product. In practice, the star-products in this article are just elegant and compact notations for the WNL form factors in the Lagrangian. They do not enter the definitions of gauge transformations that are still the standard ones with the (local) pointwise product. If it were the case, the nonassociativity would be a problem to define a deformation of the algebra, and a gauge transformation like $\psi(x) \mapsto e^{i \alpha(x)} \, \psi(x)$ under a $U(1)$ group would be ill-defined. How could we define the series expansion of the exponential without an associative multiplicative law to define an algebra? In the case of the Groenewold-Moyal product in noncommutative QFT \cite{Szabo:2001kg}, the star-product is noncommutative but associative, and one can define such a deformed algebra.}
\begin{align}
B_{\rho}(x) \star_0 B_{\sigma}(x) &\equiv B_{\rho}(x) \ast_0 B_{\sigma}(x) \\
&\equiv B_{\rho}(x) \, e^{\vartheta_0 \left( \overleftarrow{\partial_\mu} \, \eta_0^{\mu \nu} \, \overrightarrow{\partial_\nu} \right)} \, B_{\sigma}(x) \\
&= \left. e^{\vartheta_0 \left( \partial_\mu^{(i)} \, \eta_0^{\mu \nu} \, \partial_\nu^{(j)} \right)} \, B_{\rho} \left( x_i \right) \cdot B_{\sigma} \left( x_j \right) \right|_{x_i \to x_j \equiv x} \\
&= \int \dfrac{d^4p_i \, d^4p_j}{(2 \pi)^8} \, e^{-i \left( p_i + p_j \right) \cdot x \, + \, \vartheta_0 \left( - \eta_0 \, p_i \cdot p_j \right)} \, \widetilde{B}_{\rho}(p_i) \cdot \widetilde{B}_{\sigma}(p_j) \, ,
\end{align}
where $\widetilde{f}(p)$ is the Fourier transform of the function $f(x)$.

The weak isospin bosons\footnote{One introduce: (i) the structure constants $\epsilon^a_{\ bc}$ and $f^a_{\ bc}$ of the $SU(2)$ and $SU(3)$ groups, respectively; (ii) the 3 Pauli matrices $\sigma^a$ and the 8 Gell-Mann ones $\lambda^a$; (iii) the covariant derivatives acting on the adjoint representations:
\begin{align}
&\nabla_{\mu} W_{\nu} = \left(\nabla_{\mu} W_{\nu}\right)^a \frac{\sigma_a}{2} \, , \ \ 
\left(\nabla_{\mu} W_{\nu}\right)^a = \left( \delta_c^a \partial_\mu + g_2 \, \epsilon_{\ bc}^a W_{\ \mu}^b \right) W_{\ \nu}^c \, , \nonumber \\
&\nabla_{\mu} G_{\nu} = \left(\nabla_{\mu} G_{\nu}\right)^a \frac{\lambda_a}{2} \, , \ \ \left(\nabla_{\mu} G_{\nu}\right)^a = \left( \delta_c^a \partial_\mu + g_3 \, f_{\ bc}^a G_{\ \mu}^b \right) G_\nu^c \, .
\end{align}
} $W_\mu = W_\mu^a \left( \frac{\sigma^a}{2} \right)$ (resp. of gluons $G_\mu = G_\mu^a \left( \frac{\lambda^a}{2} \right)$) -- in the gauge representations $\left( \mathbf{1}, \mathbf{3} \right)_0$ and $\left( \mathbf{8}, \mathbf{1} \right)_0$, resp. -- are the gauge bosons of the group $SU(2)_W$ of coupling $g_2$ (resp. $SU(3)_C$ of coupling $g_3$). The gauge field strength tensors are
\begin{align}
&\mathcal{W}_{\mu\nu} = \mathcal{W}_{\ \mu\nu}^a \, \frac{\sigma_a}{2} \, , \ \ 
\mathcal{W}_{\ \mu\nu}^a = \partial_\mu W_{\ \nu}^a - \partial_\nu W_{\ \mu}^a + g_2 \, \epsilon_{\ bc}^a W_{\ \mu}^b W_{\ \nu}^c \, , \nonumber \\
&\mathcal{G}_{\mu\nu} = \mathcal{G}_{\ \mu\nu}^a \, \frac{\lambda_a}{2} \, , \ \ 
\mathcal{G}_{\ \mu\nu}^a = \partial_\mu G_{\ \nu}^a - \partial_\nu G_{\ \mu}^a + g_3 \, f_{\ bc}^a G_{\ \mu}^b G_{\ \nu}^c \, .
\end{align}
One defines the star-products
\begin{align}
W_{\rho}(x) \ast_w W_{\sigma}(x)
&\equiv W_{\rho}(x) \, e^{\vartheta_w \left( \overleftarrow{\partial_\mu} \, \eta_w^{\mu \nu} \, \overrightarrow{\partial_\nu} \right)} \, W_{\sigma}(x) \, , \\
W_\rho(x) \star_w W_\sigma(x)
&\equiv W_\rho(x) \, e^{\vartheta_w \left( \overleftarrow{\nabla_{\mu}} \, \eta_w^{\mu \nu} \, \overrightarrow{\nabla_{\nu}} \right)} \, W_\sigma(x) \\
&= \left. e^{\vartheta_w \left( \nabla_{\mu}^{(i)} \, \eta_w^{\mu \nu} \, \nabla_{\nu}^{(j)} \right)} \, W_\rho(x_i) \cdot W_\sigma(x_j) \right|_{x_i \to x_j \equiv x} \label{star_nabla} \\
&= \left[ \int \dfrac{d^4 p_i \, d^4 p_j}{(2 \pi)^8} e^{-i \left( p_i + p_j \right) \cdot x} \, \widetilde{W}_\rho(p_i) \cdot \widetilde{W}_\sigma(p_j) \right] \cdot \left[ \left. e^{\vartheta \left( - \eta_w \, P_{ij} \right)} \right|_{x_i \to x_j \equiv x} \right] \label{star_momentum} \\
&= W_{\rho} \ast_w W_{\sigma} + \mathcal{O}\left( \dfrac{g_2}{\Lambda_w^2} \, W^3 \right)
\end{align}
(idem for $G_\mu$ with the replacement of subscript $w \mapsto c$), where the commutators between the covariant derivatives in the adjoint representation $\nabla_\mu$ satisfy
$\left[ \nabla_{\mu}^{(i)} , \nabla_{\nu}^{(j)} \right] = \left[ \nabla_{\mu}^{(i)} , W_\nu(x_j) \right] = \left[ \nabla_{\mu}^{(i)} , \mathcal{W}_{\nu\rho}(x_j) \right] = 0$
with $x_i \neq x_j$, and\footnote{To go from Eq.~\eqref{star_nabla} to Eq.~\eqref{star_momentum}, one uses the identity
\begin{equation}
f (\partial) \, e^{-i p \cdot x} = e^{-i p \cdot x} \, f (\partial - ip) \, .
\end{equation}
}
$P_{ij} = p_i \cdot p_j + i p_j \cdot \nabla^{(i)} + i p_i \cdot \nabla^{(j)} - \nabla^{(i)} \cdot \nabla^{(j)}$ acts iteratively on the constant unit function $\mathbf{1}$ in the form factor expansion \eqref{form_factor_exp} to generate the gauge cloud.

\begin{figure}[h!]
\begin{center}
\includegraphics[height=3cm]{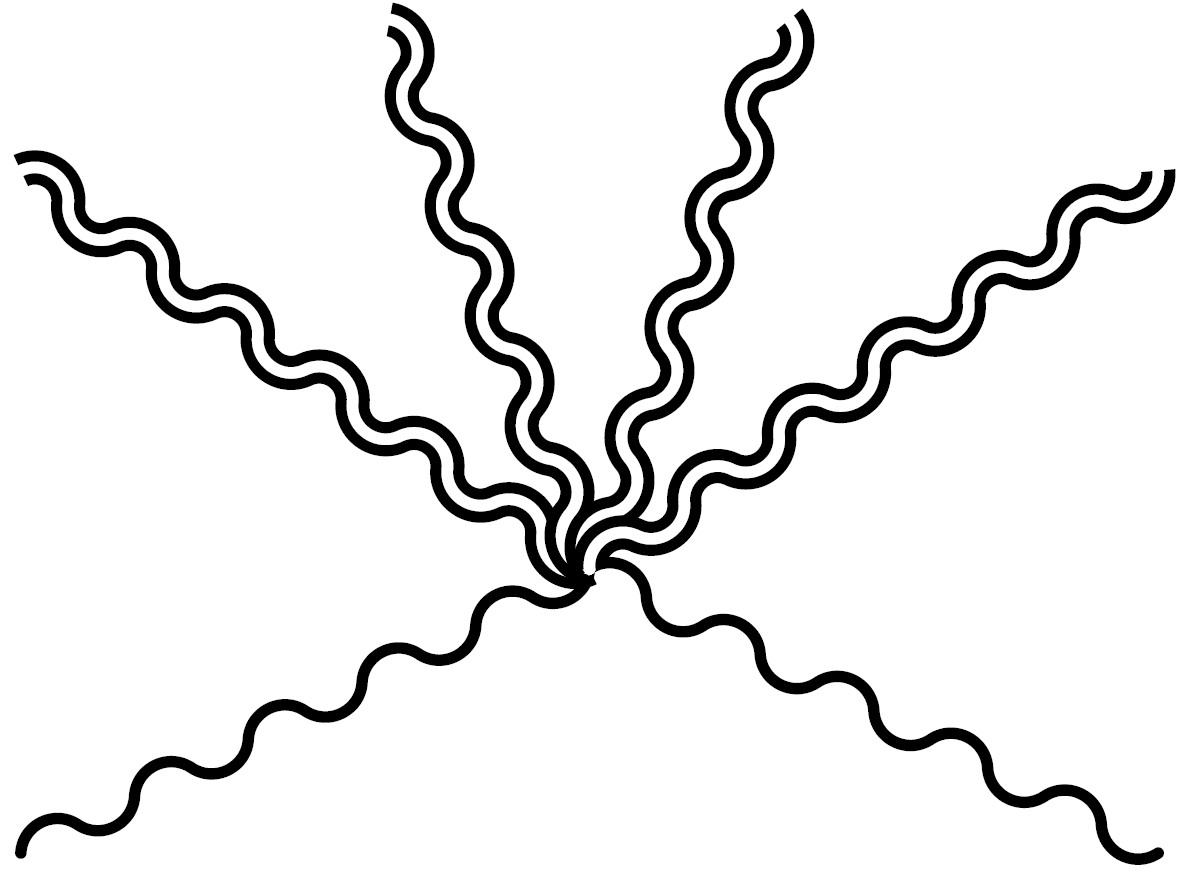}\hfill
\includegraphics[height=3cm]{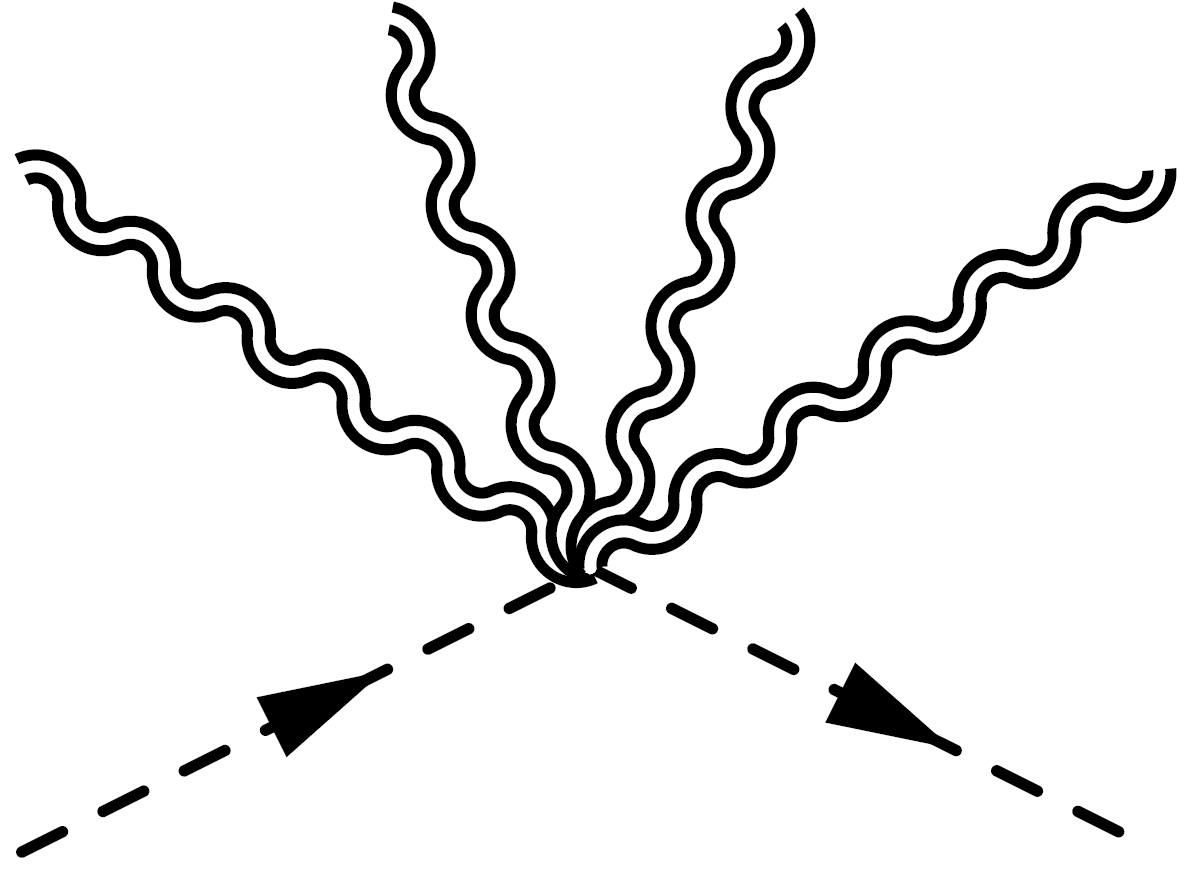}\hfill
\includegraphics[height=3cm]{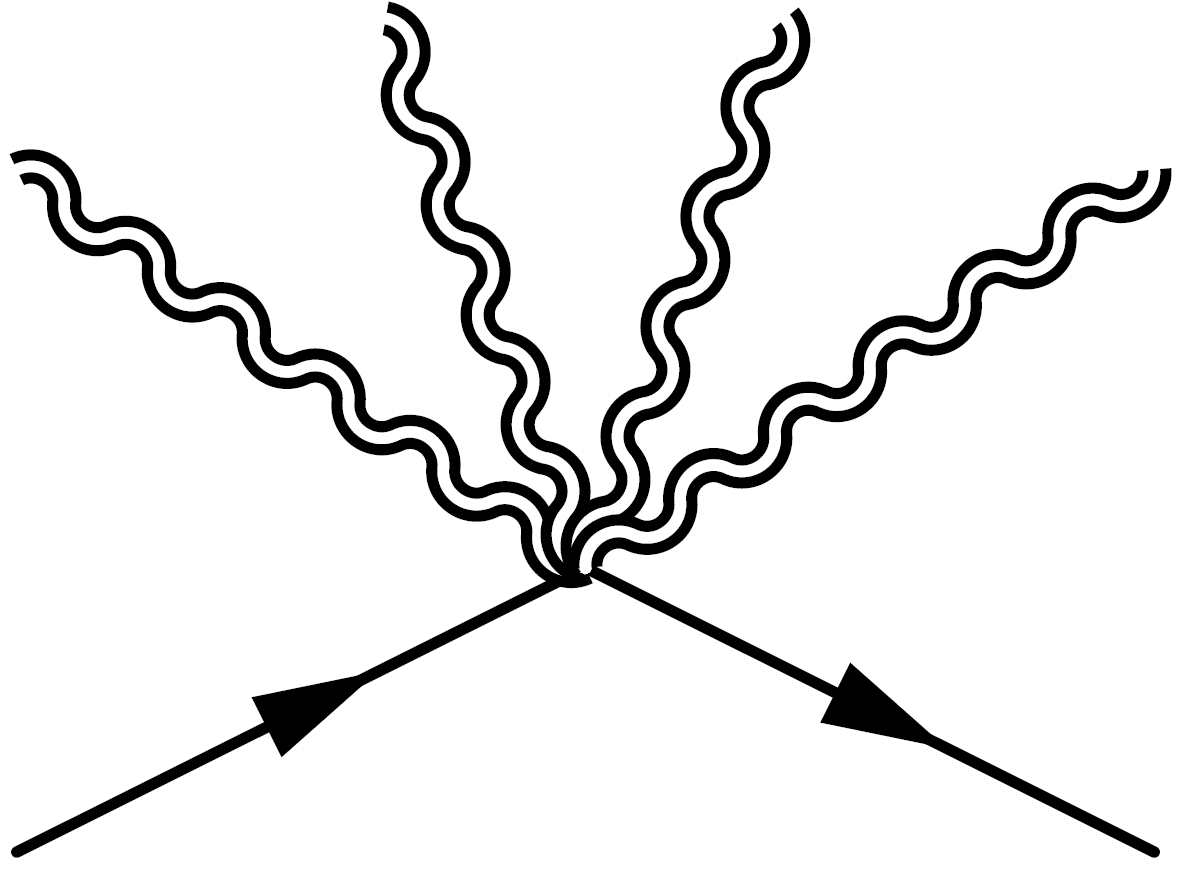}\hfill
\end{center}
\caption{\emph{Feynman diagrams of the vertex between 4 gauge bosons from the gauge clouds (double wiggly lines) and a gauge boson (left, single wiggly line), a scalar (middle, dashed line) or a fermion (right, plain line).}}
\label{particle_cloud}
\end{figure}

The noncovariant star-product $\ast_\bullet$ is thus used as a notation to ignore the perturbative expansion of the cloud of gauge bosons (\emph{gauge cloud} for short) that dresses the covariant star-product $\star_\bullet$ of fields (cf. Fig.~\ref{particle_cloud}). One can generalize these definitions to any elementary or composite bosonic field with the same gauge quantum numbers as the gauge fields, i.e. in the adjoint representation. Therefore, in this article, one attaches the star-product definitions to the gauge\footnote{The spin does not enter the definition of these star-products.} quantum numbers of the fields.
\vspace{0.2cm}

The pure gauge sector of the FSM is described by the following Lagrangian that is manifestly gauge-invariant via the Krasnikov-Terning scheme \cite{Krasnikov:1987yj, Terning:1991yt}:
\begin{align}
&- \dfrac{1}{2} \, \text{tr} \left[ \mathcal{G}_{\mu\nu} \star_c \mathcal{G}^{\mu\nu} \right]
- \dfrac{1}{2} \, \text{tr} \left[ \mathcal{W}_{\mu\nu} \star_w \mathcal{W}^{\mu\nu} \right]
- \dfrac{1}{4} \, \mathcal{B}_{\mu\nu} \star_0 \mathcal{B}^{\mu\nu} \nonumber \\
&\supseteq - \dfrac{1}{4} \, \mathcal{G}_{\ \mu\nu}^a \ast_c \mathcal{G}^{\ \mu\nu}_a
- \dfrac{1}{4} \, \mathcal{W}_{\ \mu\nu}^a \ast_w \mathcal{W}^{\ \mu\nu}_a
- \dfrac{1}{4} \, \mathcal{B}_{\mu\nu} \ast_0 \mathcal{B}^{\mu\nu} \, ,
\label{L_pure_gauge}
\end{align}
where the terms that are not displayed in the $2^\text{nd}$ line correspond to the gauge clouds that do not contribute to the tree-level propagators. Since $SU(3)_C$ is not affected by EWSB, one can already extract the gluon propagator in the Feynman-'t Hooft gauge:
\begin{equation}
\Pi_G^{ab\mu\nu}(p^2) = \dfrac{-i \, e^{- \vartheta_c \left( \eta_c \, p^2 \right)}}{p^2 + i \epsilon} \, \delta^{ab} g^{\mu\nu} \, ,
\end{equation}
which has only the canonical pole at $p^2 = 0$, aka the gluon, like in local Quantum Chromodynamics (QCD).

\subsection{Electroweak Symmetry Breaking}
\label{EWSB_section}
In the GWS model, EWSB occurs by tachyon condensation of the Higgs field, where $SU(2)_W \times U(1)_Y \rightarrow U(1)_{\text{em}}$, i.e. only the gauge symmetry $U(1)_{\text{em}}$ of Quantum Electrodynamics (QED) is linearly realized in the physical EW vacuum. In the same way as for the gauge fields, one introduces Hermitian symmetric covariant $\star_h$ and noncovariant $\ast_h$ star-products between a field in the gauge representation of the Higgs tachyon $\left( \mathbf{1}, \mathbf{2} \right)_{+1/2}$ and another field in the complex conjugate representation:
\begin{align}
H(x)^\dagger \ast_h H(x)
&\equiv H^\dagger(x) \, e^{\vartheta \left( \overleftarrow{\partial_\mu} \, \eta^{\mu \nu}_h \, \overrightarrow{\partial_\nu} \right)} \, H(x) \, ,
\label{ast_h1} \\
H^\dagger(x) \star_h H(x)
&\equiv H^\dagger(x) \, e^{\vartheta \left( \overleftarrow{\mathcal{D}_\mu} \, \eta^{\mu \nu}_h \, \overrightarrow{\mathcal{D}_\nu} \right)} \, H(x)
\label{ast_h2} \\
&= \left. e^{\vartheta \left( \overline{\mathcal{D}}_\mu^{(i)} \, \eta^{\mu \nu}_h \, \mathcal{D}_\nu^{(j)} \right)} \, H^\dagger(x_i) \cdot H(x_j) \right|_{x_i \to x_j \equiv x}
\label{ast_h3} \\
&= H^\dagger(x) \ast_h H(x)  + \mathcal{O}\left( \dfrac{g_1}{\Lambda_h^2} \, H^\dagger H B \right) + \mathcal{O}\left( \dfrac{g_2}{\Lambda_h^2} \, H^\dagger H W \right) \, , \label{ast_h4}
\end{align}
where the commutator between covariant derivatives in the (anti)-fundamental representations\footnote{The action of these covariant derivatives reads
\begin{equation}
\mathcal{D}_\mu H = \left( \partial_\mu -i g_2 \, W_{\mu} -i \frac{g_1}{2} \, B_{\mu} \right) \cdot H \, ,
\ \ \overline{\mathcal{D}}_\mu H^\dagger = \left( \mathcal{D}_\mu H \right)^\dagger \, .
\end{equation}
}
$\mathcal{D}_\mu$ and $\overline{\mathcal{D}}_\mu$ satisfies $\left[ \overline{\mathcal{D}}_\mu^{(i)}, \mathcal{D}_\nu^{(j)} \right] = 0$ with $x_i \neq x_j$, and the gauge clouds (cf. Fig.~\ref{particle_cloud}) are in the ellipses $\mathcal{O}(\cdots)$.

\begin{figure}[h!]
\begin{center}
\includegraphics[height=3cm]{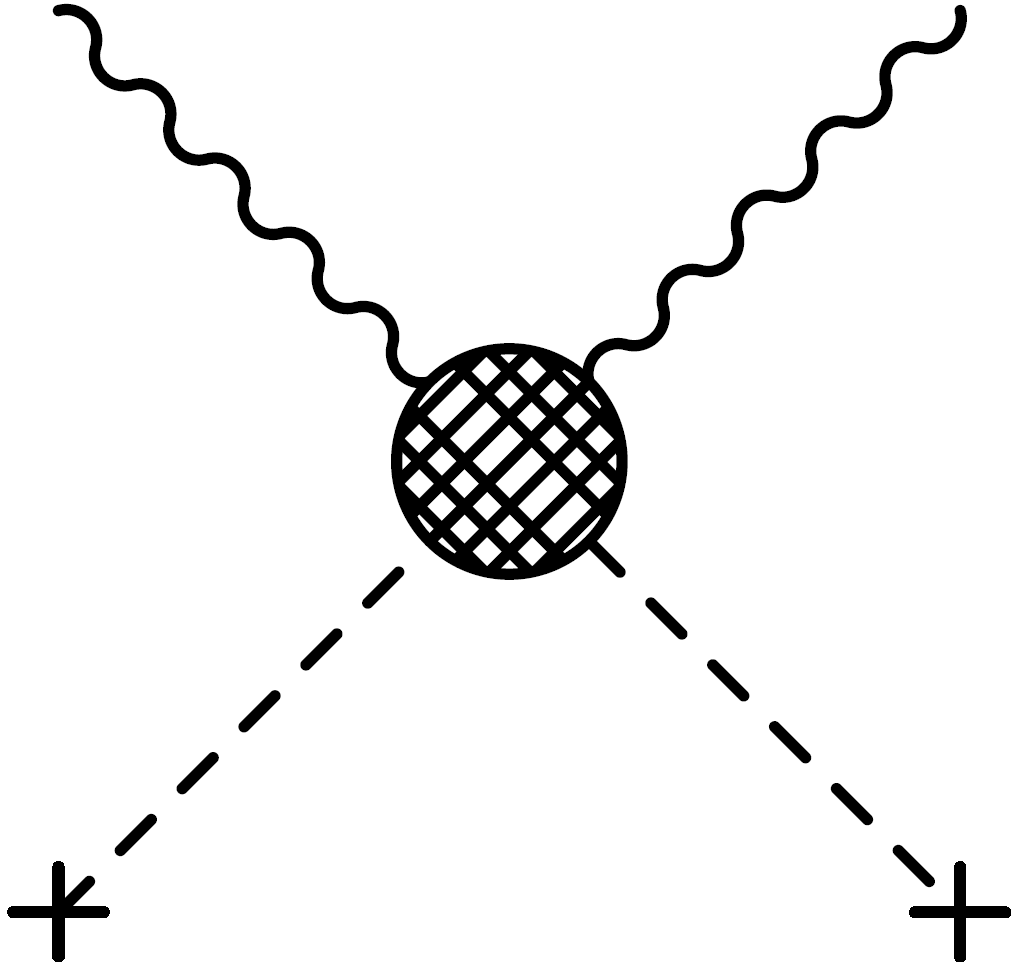}\hfill
\includegraphics[height=3cm]{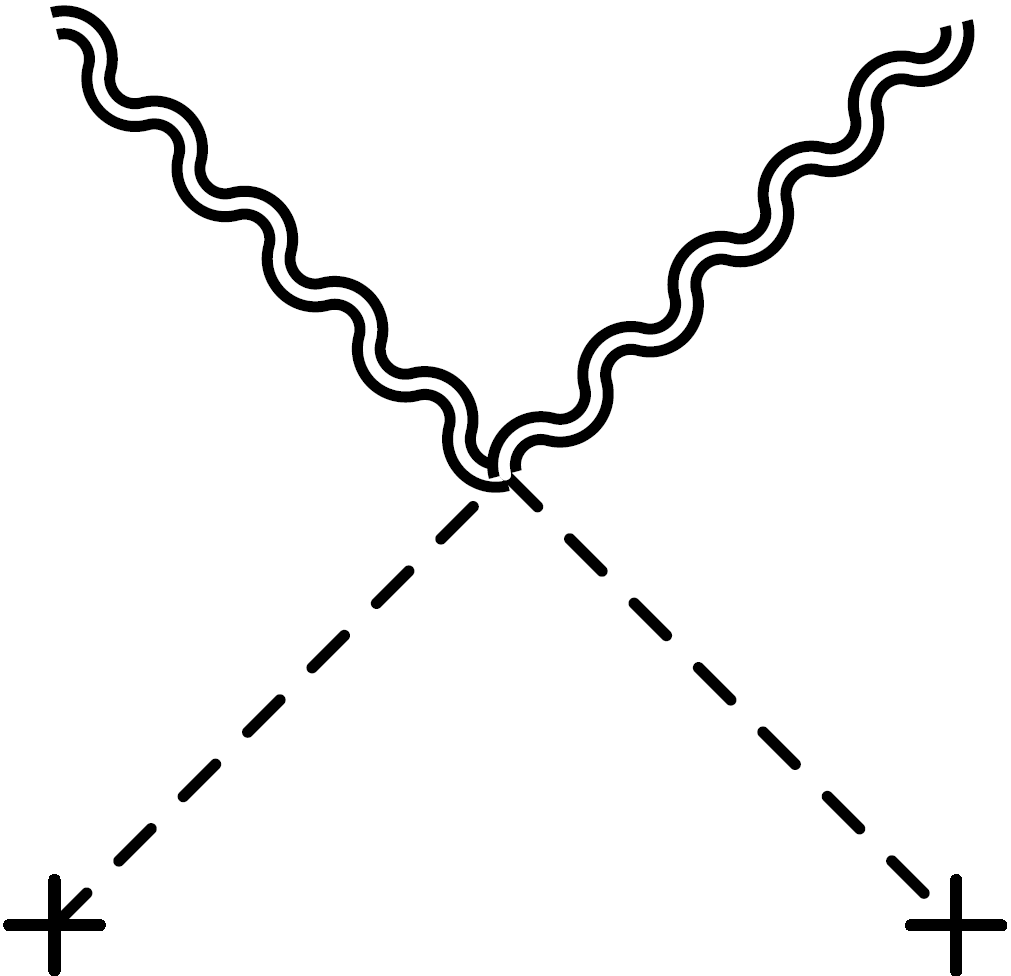}\hfill
\end{center}
\caption{\emph{Feynman diagrams in unitary gauge, where the blob is a WNL vertex, crosses are insertions of Higgs vev's, and single (doubly) wiggly lines are gauge bosons from outside (inside) the gauge clouds. Left: the covariant derivatives in the quadratic term in Higgs tachyons \eqref{Higgs_kin} implies an interaction between a gauge boson with 2 insertions of Higgs vev's, giving a WNL contribution to the mass term of the gauge boson. Right: when the Higgs potential involves a gauge cloud, one gets a tower of local interactions between a gauge boson with 2 insertions of Higgs vev's, giving a new WNL contribution to the quadratic terms in gauge bosons.}}
\label{vev_cloud}
\end{figure}

Concerning the vev $v$ of $H(x)$, one has the following important properties:
\begin{equation}
v \ast_h H(x) = v \cdot H(x) \ \ \ 
\text{and} \ \ \ 
v \star_h v = v^2 + \mathcal{O}(W^2) + \mathcal{O}(B^2) \, ,
\label{vev_properties}
\end{equation}
where the $2^\text{nd}$ one means that if a term like $H(x)^\dagger \star_h H(x)$ appears in the Higgs potential, it gives an infinite-tower of higher-derivative corrections to the Lagrangian quadratic terms of the weak gauge bosons (cf. Fig.~\ref{vev_cloud}). This crucial property spoils the ghost-free factorization of the propagators, thus one should use only the $\star_0$-product in the Higgs potential\footnote{This point is corrected in the \texttt{arXiv} version (\texttt{v7}) of Ref.~\cite{Nortier:2023dkq}. Of course, one can only use $\star_0$ between gauge-singlet composite fields, like $H^\dagger \cdot H$, to not spoil gauge-invariance. This fixes completely the WNL deformation of the quartic Higgs potential.}.

One can then build the following Lagrangian for the kinetic and potential terms of the Higgs tachyon $(v, \ \lambda_h > 0)$:
\begin{align}
\mathcal{L}_{\text{H}} &= \mathcal{D}_\mu H^\dagger \star_h \mathcal{D}^\mu H - V(H, H^\dagger) \, ,
\label{Higgs_Lag_1}
\end{align}
\begin{equation}
V(H, H^\dagger) = -\mu^2 \, H^\dagger \cdot H + \lambda_h \left( H^\dagger \cdot H \right) \star_0 \left( H^\dagger \cdot H \right) \, ,
\label{Higgs_Lag_2}
\end{equation}
with a WNL quartic self-coupling\footnote{One could also introduce a local quartic coupling $\propto \left( H^\dagger \cdot H \right)^2$, but it is shown in Ref.~\cite{Nortier:2023dkq} (\texttt{v7}) that its coefficient must vanish to get a ghost-free (Higgs) $H$-boson propagator at tree-level.}. Then, one performs the replacement $\left( \star_h \mapsto \ast_h \right)$ in Eqs.~\eqref{Higgs_Lag_1} and \eqref{Higgs_Lag_2}, i.e. one drops the gauge cloud from the covariant star-products of Higgs tachyons. Indeed, this gauge cloud does not contribute to the tree-level propagators after EWSB by using the properties \eqref{vev_properties}.

Now, consider the polar representation of the Higgs tachyon around its vev as
\begin{equation}
H = e^{i \, \frac{\pi^a(x)}{v} \, \sigma_a} \, H_U \, , \ \ \ 
H_U(x) = \dfrac{1}{\sqrt{2}}
\begin{pmatrix}
0 \\
v + h(x)
\end{pmatrix} \, , \ \ \ 
v = \sqrt{\dfrac{\mu^2}{\lambda}} \, ,
\end{equation}
where $h(x)$ is the radial mode, aka the Higgs $H$-boson. In order to study the physical spectrum, one chooses the unitary gauge, where the pion fields $\pi^a(x)$ are eaten by the $W_\mu$ and $B_\mu$ gauge bosons, such that $H(x) \equiv H_U(x)$. One can then extract the vector boson masses from the quadratic term in Higgs tachyons with covariant derivatives
\begin{align}
\mathcal{D}_\mu H_U^\dagger \ast_h \mathcal{D}^\mu H_U
&= \dfrac{1}{2} \left[
\begin{pmatrix}
0 & v
\end{pmatrix}
\cdot \left( \dfrac{g_2}{2} \, W_{\mu}^a \sigma_a + \dfrac{g_1}{2} B_{\mu} \right)
\right] \ast_h \left[
\left( \dfrac{g_2}{2} \, W_{\mu}^b \sigma_b + \dfrac{g_1}{2} B^{\mu} \right) \cdot
\begin{pmatrix}
0 \\
v
\end{pmatrix}
\right] \nonumber \\
&= \dfrac{v^2}{8} \left[ g_2^2 \left( W_\mu^1 \ast_h W^\mu_1
+ W_\mu^2 \ast_h W^\mu_2 \right) \right. \nonumber \\
&+ \left. \left( -g_2 \, W_\mu^3 + g_1 \, B_\mu \right)
\ast_h \left( -g_2 \, W^\mu_3 + g_1 \, B^\mu \right) \right] .
\label{Higgs_kin}
\end{align}
These mass terms mix the $W_\mu^3$ and $B_\mu$ fields, and one must ensure that they combine with the kinetic terms \eqref{L_pure_gauge} to give propagators with ghost-free factorization like, for the photon,
\begin{equation}
\Pi_A^{\mu\nu}(p^2) = \dfrac{-i \, e^{- \vartheta_0 \left( \eta_0 \, p^2 \right)}}{p^2 + i \epsilon} \, g^{\mu\nu} \, ,
\label{propa_A}
\end{equation}
and for the weak bosons of mass $M_V$,
\begin{equation}
\Pi_V^{\mu\nu}(p^2) = \dfrac{-i \, e^{- \vartheta_w \left( \eta_w \, p^2 \right)}}{p^2 - M_V^2 + i \epsilon} \left( g^{\mu\nu} - \dfrac{p^\mu p^\nu}{M_V^2} \right) .
\label{propa_V}
\end{equation}
For this to happen, one can see from Eq.~\eqref{Higgs_kin} that the involved noncovariant star-products must be the same:
\begin{equation}
\ast_w \equiv \ast_0 \equiv \ast_h \ \ 
\Rightarrow \ \ 
\eta_w = \eta_0 = \eta_h \, , \ \ 
\vartheta_w(z) = \vartheta_0(z) = \vartheta_h(z) \, ,
\end{equation}
otherwise, the WNL form factors do not factorize between the kinetic and mass terms, and the ghost-free condition is spoiled. Physically, since the longitudinal polarizations of the weak bosons come from the Higgs field, it is thus not surprising that a ghost-free factorization of the propagator in unitary gauge requires the same WNL form factors to occur.

Now, one can perform a rotation in the basis of the physical vector bosons:
\begin{align}
&W_\mu^\pm = \dfrac{1}{\sqrt{2}} \left( W_\mu^1 \mp W_\mu^2 \right) \, ,
\ \ \ 
\begin{pmatrix}
Z_\mu \\
A_\mu
\end{pmatrix}
=
\begin{pmatrix}
\cos \theta_w & -\sin \theta_w \\
\sin \theta_w & \cos \theta_w
\end{pmatrix}
\begin{pmatrix}
W^3_\mu \\
B_\mu
\end{pmatrix} \, .
\end{align}
The weak mixing angle $\theta_w$ and the $U(1)_{\text{em}}$ gauge coupling $g_e$ are defined as usual:
\begin{align}
\tan \theta_w = \dfrac{g_1}{g_2} \, , \ \ \ 
g_e = g_2 \, \sin \theta_w \, .
\end{align}
The quadratic Lagrangian of the physical vector bosons reads
\begin{align}
&- \dfrac{1}{4} \, \mathcal{A}_{\mu\nu} \ast_w \mathcal{A}^{\mu\nu}
- \dfrac{1}{2} \, \mathcal{W}_{\mu\nu}^+ \ast_w \mathcal{W}^{-\mu\nu}
- \dfrac{1}{4} \, \mathcal{Z}_{\mu\nu} \ast_w \mathcal{Z}^{\mu\nu}
\nonumber \\
&+ \dfrac{M_A^2}{2} \, A_\mu \ast_w A^\mu
+ M_W^2 \, W^+_{\mu} \ast_w W^{-\mu}
+ \dfrac{M_Z^2}{2} \, Z_\mu \ast_w Z^\mu \, , \\
&\text{with} \
M_A=0 \, , \ \ 
M_W = g_2 \, \dfrac{v}{2} \, , \ 
\ \ \text{and} \ \ 
M_Z = \dfrac{M_W}{\cos \theta_w} \, .
\end{align}
with the field strength tensors
\begin{equation}
\mathcal{A}_{\mu\nu} = \partial_\mu A_\nu - \partial_\nu A_\mu \, , \ \ 
\mathcal{W}_{\mu\nu}^\pm = \partial_\mu W^\pm_{\nu} - \partial_\nu W^\pm_{\mu} \, , \ \ 
\mathcal{Z}_{\mu\nu} = \partial_\mu Z_\nu - \partial_\nu Z_\mu \, .
\end{equation}
From this quadratic Lagrangian, one gets the propagators for the photon $A_\mu$, and the weak bosons $V_\mu = Z_\mu$ or $W_\mu^\pm$, in Feynman and unitary gauges, respectively: Eqs.~\eqref{propa_A} and \eqref{propa_V}. Concerning the $H$-boson $h(x)$, its quadratic Lagrangian is
\begin{align}
&\dfrac{1}{2} \, \partial_\mu h \ast_w \partial^\mu h - \dfrac{M_H^2}{2} \, h \ast_w h \, , \\
&\text{with} \ M_H = \sqrt{2} \mu = \sqrt{2 \lambda_h} v \, ,
\label{h_quad}
\end{align}
and one gets the ghost-free propagator
\begin{equation}
\Pi_H(p^2) = \dfrac{i \, e^{- \vartheta_w \left( \eta_w \, p^2 \right)}}{p^2 - M_H^2 + i \epsilon} \, .
\label{H-propa}
\end{equation}

This analysis shows that there is no ghost at tree-level in the bosonic sector of the proposed FSM. As discussed in detail in Ref.~\cite{Nortier:2023dkq}, the main difference with the usual string-inspired models of the literature \cite{Gama:2018cda, Hashi:2018kag, Koshelev:2020fok, Nortier:2023dkq} is that \emph{both} the kinetic term and the potential of the Higgs field are smeared via the star-products. Instead, in string-inspired models, only the fields in \emph{either} the quadratic \emph{or} the self-interaction terms are smeared, but not in both.

A crucial observation for the phenomenology of the FSM is that the tree-level expressions of the particle masses after EWSB are the same as in the SM in terms of the Lagrangian parameters. This means that the difference between the perturbative spectra of the 2 models arises only at loop-level. This is in contrast with what happens when tachyon condensation is realized in the string-inspired formalism with smeared fields \cite{Gama:2018cda, Hashi:2018kag, Nortier:2023dkq}, e.g. Ref.~\cite{Krasnikov:2022xsi} modifies the tree-level predictions of the weak boson masses via a contribution of a WNL scale, while the model is not ghost-free for the same reasons explained in Refs.~\cite{Hashi:2018kag, Nortier:2023dkq}. 

Because fuzziness does not change the SM tree-level predictions for the perturbative mass spectrum, the parameter
\begin{equation}
\rho = \dfrac{M_W^2}{M_Z^2 \, \cos^2 \theta_w} 
\end{equation}
is equal to 1 in both the SM and the FSM at tree-level, which is not trivial when one modifies how EWSB occurs. This issue is related to the SM approximate global $SO(4)$ symmetry of the Higgs potential, which is spontaneously broken by EWSB to the $SO(3)$ custodial symmetry \cite{Sikivie:1980hm} that implies $\rho = 1$ at tree-level. In the SM, these global symmetries become exact when the Yukawa couplings and $g_1 \to 0$. In the FSM, the Higgs potential \eqref{Higgs_Lag_2} still involves only the $SO(4)$ invariant combination $H^\dagger \cdot H$, and the star-product $\star_0$ does not break it. The sources of $SO(3)$ breaking are still the same operators as in the SM but with a WNL deformation. After a truncated expansion of the form factors \eqref{form_factor_exp}, this means from a SMEFT perspective \cite{Falkowski:2023hsg} that the additional contributions from WNL to the $SO(3)$ breaking come from operators that scale as $\eta_\bullet^{n} \, g_{SM}$, where $g_{SM}$ is some SM coupling that break the custodial symmetry, $\eta_\bullet$ is some fuzzy-plaquette, and $n \in \mathbb{N}^*$ depends on $\vartheta_\bullet(z)$. Therefore, the WNL contributions are expected to be subdominant with respect to the SM ones. A detailed analysis of the constraints on the fuzzy-plaquettes coming from EW precision tests would be interesting for future works.

\section{Fermionic Fields}
\label{fermionic_fields}

\subsection{Dirac \& Weyl Fermions}
Since the fundamental building blocks of the SM fermions are Weyl spinors, where different chiralities\footnote{One considers the Dirac $\gamma$-matrices in the chiral representation. We work in the 4-component formalism where, for a spinor $\Psi$, the chiral decomposition $\Psi = \Psi_L + \Psi_R$ is given in terms of the chiral projectors $P_{L/R}$ by $\Psi_{L/R} = P_{L/R} \, \Psi$.} belong to different representations of the EW gauge group, it is tempting to define different star-products for different chiralities, where the kinetic terms would be
\begin{equation}
\dfrac{i}{2} \left( \overline{\Psi}_{L} \star_{L} \slashed{\partial} \Psi_{L}
+ \overline{\Psi}_{R} \star_{R} \slashed{\partial} \Psi_{R} \right) + \text{H.c.} \, ,
\end{equation}
with $\slashed{\partial} = \gamma^\mu \partial_\mu$.
Nevertheless, mass terms mix chiralities, which imposes the same expression for the noncovariant star-products:
\begin{equation}
- M_\psi \left( \overline{\Psi}_L \ast_\psi \Psi_R + \overline{\Psi}_R \ast_\psi \Psi_L \right)
\ \ \ \Rightarrow \ \ \ 
\ast_\psi \equiv \ast_L \equiv \ast_R \, ,
\end{equation}
in order to get a ghost-free propagator for a Dirac fermion of the form
\begin{equation}
\Pi_\psi(p) = \dfrac{i \, e^{- \vartheta_\psi \left( \eta_\psi \, p^2 \right)}}{\gamma \cdot p - M_\psi + i \epsilon} \, .
\label{propa_fermion}
\end{equation}

In the same way as for the bosons, one defines the covariant $\star_\psi$ and noncovariant $\ast_\psi$ star-products between 2 elementary and/or composite fermion fields (in the complex conjugate gauge representation of each other) by performing the following substitutions in Eqs.~\eqref{ast_h1}-\eqref{ast_h4}: $H \mapsto \Psi$, $H^\dagger \mapsto \overline{\Psi} = \Psi^\dagger \gamma^0$, the subscript $h \mapsto \psi$, and the covariant derivative $\mathcal{D}_\mu$ acting on the appropriate fermion representation.

Now, one introduces the 3 generations of SM fermion fields: the left-handed quarks $Q_L^i = \left( u_L^i, d_L^i \right)^T \in \left( \mathbf{3}, \mathbf{2} \right)_{+1/6}$, the right-handed $d$-type quarks $d_R^i \in \left( \mathbf{3}, \mathbf{1} \right)_{-1/3}$, the right-handed $u$-type quarks $u^i_R \in \left( \mathbf{3}, \mathbf{1} \right)_{+2/3}$, the left-handed leptons $L_L^i = \left( \nu_L^i, e_L^i \right)^T \in \left( \mathbf{1}, \mathbf{2} \right)_{-1/2}$, and the right-handed $e$-type leptons $e_R^i \in \left( \mathbf{1}, \mathbf{1} \right)_{-1}$.
The gauge-invariant kinetic terms for these fermions are
\begin{align}
\dfrac{i \, \delta_{ij}}{2} \left[
\overline{Q}_{L}^i \star_q \left( \slashed{\mathcal{D}} Q_{L}^j \right) 
+ \overline{d}_{R}^i \star_q \left( \slashed{\mathcal{D}} d_{R}^j \right)
+ \overline{u}_{R}^i \star_q \left( \slashed{\mathcal{D}} u_{R}^j \right)
+ \overline{L}_{L}^i \star_\ell \left( \slashed{\mathcal{D}} L_{L}^j \right)
+ \overline{e}_{R}^i \star_\ell \left( \slashed{\mathcal{D}} e_{R}^j \right)
\right] + \text{H.c.} \, ,
\end{align}
where one has a star-product $\star_q$ for quarks and another one $\star_\ell$ for leptons. We comment on the choice of the same star-products for quarks and leptons in different generations in the following section.

\subsection{Yukawa Couplings \& Flavor Mixing}
In order to write gauge-invariant Yukawa terms, with the covariant star-products between chiral fermions in different representations of the EW gauge group, one can use the composite fields made of 1 chiral fermion and 1 Higgs tachyon (built with the usual pointwise product). For 1 generation, the Yukawa operators (with positive couplings) would be\footnote{We do not address the issue of neutrino masses in this article.}
\begin{align}
\mathcal{L}_{\text{Y}}^{(1)} &=
- \lambda_d \left( \overline{Q}_L \cdot H \right) \star_q d_R
- \lambda_u \left( \overline{Q}_{L} \cdot \overline{H} \right) \star_q u_R \nonumber \\
&- \lambda_e \left( \overline{L}_L \cdot H \right) \star_\ell e_R
+ \text{H.c.} \, , \ \ \ \overline{H} = i \sigma_2 \cdot H^* \, , \\
&\supseteq
- m_d \, \overline{d}_L \ast_q d_R
- m_u \, \overline{u}_L \ast_q u_R
- m_e \, \overline{e}_L \ast_\ell e_R
+ \text{H.c.} \, , \\
&\text{with} \ \ \ 
m_d = \dfrac{\lambda_d v}{\sqrt{2}} \, , \ \ \ 
m_u = \dfrac{\lambda_u v}{\sqrt{2}} \, , \ \ \ 
m_e = \dfrac{\lambda_e v}{\sqrt{2}} \, , \ \ \ 
\end{align}
where the mass terms of the $2^\text{nd}$ line are extracted from the quadratic terms involving the noncovariant star-products (like for the boson fields). Again, the tree-level spectrum is ghost-free and is the same as in the SM with 1 generation.

The generalization to 3 generations with complex Yukawa couplings is straightforward:
\begin{align}
\mathcal{L}_{\text{Y}}^{(3)} &=
- \lambda_d^{ij} \left( \overline{Q}_L^i \cdot H \right) \star_q d_R^j
- \lambda_u^{ij} \left( \overline{Q}_{L}^i \cdot \overline{H} \right) \star_q u_R^j
- \lambda_e^{ij} \left( \overline{L}_L^i \cdot H \right) \star_\ell e_R^j
+ \text{H.c.} \, .
\end{align}
We stress that it is necessary to use the same covariant star-products between different generations that belong to the same gauge representation. Indeed, the WNL form factors in the quadratic terms need to factorize to keep ghost-free propagators \eqref{propa_fermion}. It follows that both the covariant and noncovariant star-products are linear in flavor space, so the flavor rotations:
\begin{align}
&\text{Quarks:} \ \ \ 
d_{L/R}^i \mapsto D_{L/R}^{ij} \, d_{L/R}^j \, ,
\ \ \ 
u_{L/R}^i \mapsto U_{L/R}^{ij} \, u_{L/R}^j \, , \\
&\text{Leptons:} \ \ \ 
e_{L/R}^i \mapsto E_{L/R}^{ij} \, e_{L/R}^j \, ,
\ \ \ 
\nu_L^i \mapsto N_L^{ij} \, \nu_L^j \, ,
\end{align}
can go through the star-products without complications. By using the usual field redefinition procedure, e.g. Ref.~\cite{Halzen:1984mc}, it is thus straightforward to show that the FSM has the same qualitative features as the SM concerning flavor mixing, which are encoded in the Cabibbo-Kobayashi-Maskawa (CKM) matrix \cite{Cabibbo:1963yz, Kobayashi:1973fv}: there is only 1 physical $CP$-violating phase coming from the Yukawa couplings (with 3 fermion generations) and no flavor-changing neutral currents (FCNC's) via the Glashow–Iliopoulos–Maiani (GIM) mechanism\footnote{This result holds because we have defined a minimal WNL deformation of the (local) SM Yukawa operators: only 1 WNL operator is associated to a local counterpart. There is thus the same number of Yukawa couplings in the FSM as in the SM. If we also defined star-products between non-complex conjugate gauge representations, we would get operators like $\left(\overline{L}^i_L \star_{\ell h} H \right) \star_\ell e_R^j$, thus more Yukawa couplings than in the SM with their associated issues (other physical phases, FCNC's).} \cite{Glashow:1970gm}.

In the limit of vanishing Yukawa couplings, this minimal FSM exhibits the same flavor symmetry group as the SM:
\begin{equation}
U(3)_Q \times U(3)_u \times U(3)_d \times U(3)_L \times U(3)_e \, .
\end{equation}
Since the star-products introduce a WNL deformation of the local Yukawa operators, the breaking effects are also controlled by the same numerical values of the Yukawa couplings, with subleading contributions coming from higher-dimensional operators suppressed by the WNL scales, similarly to the discussion on the $SO(3)$ custodial symmetry in Section~\ref{EWSB_section}. Therefore, the FSM offers a framework where the contributions to the breaking of flavor symmetries are naturally suppressed, for the same reason they are suppressed in the SM. A corollary is that the baryon and lepton numbers are also perturbatively conserved due to an accidental $U(1)_B \times U(1)_L$ global symmetry of the Lagrangian.

\section{Classicalization via Fuzziness?}
In order to illustrate the effect of fuzziness on the UV-behavior of scattering amplitudes, we focus on a toy model made of the pure Higgs sector of the FSM, i.e. one takes the limit of vanishing gauge and Yukawa interactions for the $H$-boson, and we perform the academic exercise of computing the tree-level scattering amplitude $\mathcal{M}_{22}$ of the process
\begin{equation}
H \left( p_1 \right) + H \left( p_2 \right)
\rightarrow H \left( p_3 \right) + H \left( p_4 \right) \, .
\end{equation}
We introduce the Mandelstam variables $s = \left( p_1 + p_2 \right)^2$, $t = \left( p_1 - p_3 \right)^2$ and $u = \left( p_1 - p_4 \right)^2$ with $s+t+u=4 M_H^2$.

The quadratic terms for the $H$-boson Lagrangian are given in Eq.~\eqref{h_quad}, and the self-interaction terms read
\begin{equation}
- \lambda_h v \, h \ast_w h^2 - \dfrac{\lambda_h}{4} \, h^2 \ast_w h^2 \, .
\end{equation}
The propagator is given in Eq.~\eqref{H-propa}, and the Feynman rules for the cubic and quartic self-interactions are
\begin{align}
&-i \lambda_h v \sum_{\sigma \in S_3} e^{\vartheta_w \left[- \eta_w \, p_{\sigma(1)} \cdot \left( p_{\sigma(2)} + p_{\sigma(3)} \right) \right]} \, ,\\
&- \dfrac{i\lambda_h}{4} \sum_{\sigma \in S_4} e^{\vartheta_w \left[- \eta_w  \left( p_{\sigma(1)} + p_{\sigma(2)} \right) \cdot \left( p_{\sigma(3)} + p_{\sigma(4)} \right) \right]} \, ,
\end{align}
respectively, where $S_n$ is the set of permutations $\sigma$ of $n \in \mathbb{N}^*$ elements, and all momenta are chosen in-going the vertices.

At tree-level, the cubic self-coupling generates 3 diagrams (the $stu$-channels), and the quartic one gives 1 contact diagram. The scattering amplitude can be expressed as $\mathcal{M}_{22}(s,t) = \mathcal{M}_c(s) + \mathcal{M}_c(t) + \mathcal{M}_c(u)$, with
\begin{align}
\mathcal{M}_c \left( q^2 \right) &= -2 \lambda_h \, e^{-2 \vartheta_w \left( \eta_w M_H^2 \right)-\vartheta_w \left( \eta_w q^2 \right)}
\left[\frac{2 \lambda_h v^2 \left(2 \, e^{\vartheta_w \left(\eta_w M_H^2 \right)}+e^{\vartheta_w \left( \eta_w  q^2 \right)}\right)^2}{q^2-M_H^2}+e^{2 \vartheta \left( \eta_w  q^2 \right)}\right] .
\end{align}
To be concrete, one can choose exponential form factors with $\vartheta_w(z) = (-z)^p$, and $p \in \mathbb{N}^*$, where the propagator \eqref{H-propa} is UV-damped for Euclidean momenta $p_E^2 \gg \Lambda_w^2$. For $p=1$, in the hard scattering limit ($s \to + \infty$, $t \to - \infty$ and $s/t$ fixed):
\begin{equation}
\mathcal{M}_{22}(s,t) \sim - 16 \lambda_h^2 v^2 \, \dfrac{e^{\eta_w  s}}{s} \, ,
\end{equation}
so the amplitude blows up exponentially in the deep-UV\footnote{This is known in string-inspired models \cite{Pius:2016jsl, Carone:2016eyp, Buoninfante:2018mre, Briscese:2018oyx, Chin:2018puw, Pius:2018crk, DeLacroix:2018arq, Briscese:2021mob, Koshelev:2021orf, Buoninfante:2022krn, Mo:2022szw, Buoninfante:2023dyd} for odd $p$. However, in our case, because of the competition between the form factors in propagators and vertices, taking an even $p$ does not improve the situation.}. Therefore, \emph{perturbative} unitarity is lost for $\sqrt{s} \gg \Lambda_w$, but not necessarily at the \emph{nonperturbative} level.

\begin{figure}[h!]
\begin{center}
\includegraphics[height=3cm]{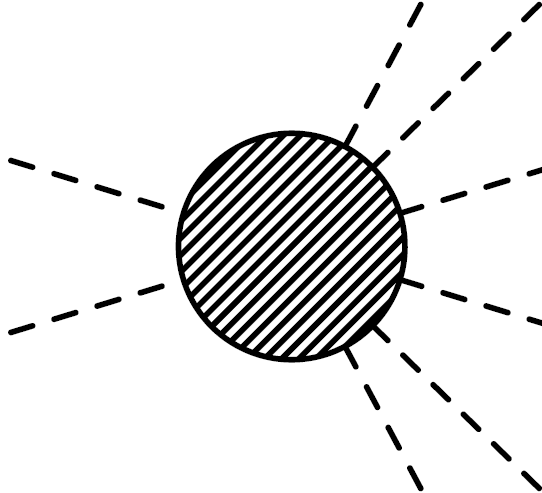}
\end{center}
\caption{\emph{Sketch of the scattering process: $H + H \rightarrow \text{Higgsion} \rightarrow H + H + \cdots + H$. The Higgsion is represented as a shaded blob, and the $H$-bosons as dashed lines.}}
\label{Higgsion}
\end{figure}

To understand what could save unitarity at the nonperturbative level, one performs the following field redefinition:
\begin{equation}
\Phi(x) = e^{\frac{\eta_w}{2} \square} \, H(x)
\end{equation}
in the Lagrangian of the Higgs tachyon \eqref{Higgs_Lag_1}-\eqref{Higgs_Lag_2}. Then, one truncates the low-energy expansion of the form factors \eqref{form_factor_exp} and, at the end, one canonically normalizes the $\Phi$-kinetic term. The local effective Lagrangian is
\begin{align}
\partial_\mu \Phi^\dagger \cdot \partial^\mu \Phi + \mu^{\prime 2} \, \Phi^\dagger \cdot \Phi - \lambda_h^{\prime 2} \left( \Phi^\dagger \cdot \Phi \right)^2 - 2 \lambda_h \eta_w \left( \Phi^\dagger \cdot \Phi \right) \cdot \left( \partial_\mu \Phi^\dagger \cdot \partial^\mu \Phi \right) + \mathcal{O} \left( \eta_w^2 \right) \, ,
\end{align}
with $\mu^{\prime 2} = \mu^2 \left( 1 - \eta_w \mu^2 \right) > 0$ and $\lambda_h^{\prime 2} = \lambda_h \left( 1-2\eta_w \mu^2 \right)>0$. The last term is a particular nonrenormalizable operator of the type $\left( \Phi^\dagger \Phi \right) J(s)$, with an energy source $J = \partial_\mu \Phi^\dagger \partial^\mu \Phi$ growing with $s$. Such operator has been proposed to restore unitarity via nonperturbative effects in Ref.~\cite{Dvali:2010jz}, offering a non-Wilsonian UV-completion, i.e. without introducing new degrees of freedom. In this scenario, for $\sqrt{s} \gg \Lambda_w$, the $2 \to 2$ hard scattering is actually exponentially suppressed by the formation of a classical $\Phi$-configuration (classicalon) that decays preferentially into $N \gg 1$ soft-quanta (cf. Fig~\ref{Higgsion}), and restores unitarity\footnote{We stress that such non-Wilsonian UV-completion does not rely on the perturbative renormalization program or the existence of an interacting UV-fixed point at the nonperturbative level. Deep-UV processes are converted to deep-IR ones via the production of many soft quanta.}.\newpage

\begin{figure}[h!]
\begin{center}
\includegraphics[width=13.9cm]{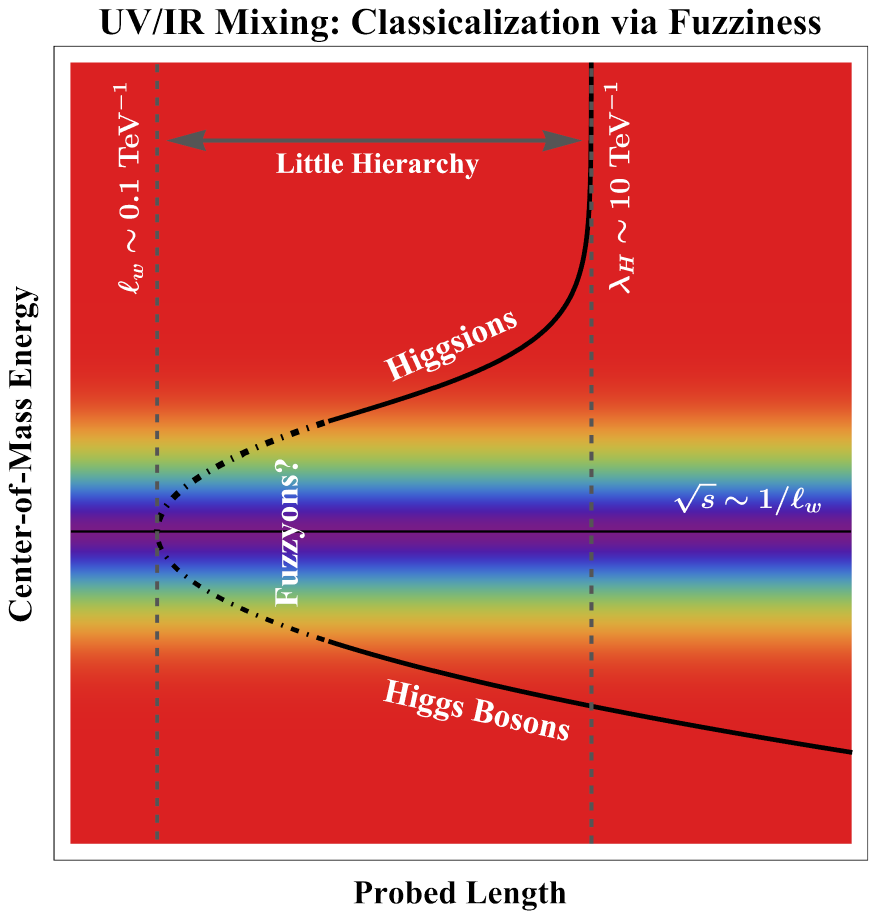}
\end{center}
\caption{\emph{Sketch of UV/IR mixing by classicalization of the Higgs sector proposed in Ref.~\cite{Dvali:2010jz}. In a hard scattering experiment with $\sqrt{s} \ll \Lambda_w = 1/\ell_w$, one probes shorter distances by increasing $\sqrt{s}$ as in a local QFT. Higgs bosons are produced via perturbative processes, like at the LHC. However, for $\sqrt{s} \gg \Lambda_w$, the Vainshtein radius is $R_V \gg \ell_w$, and one enters a semiclassical regime where $N \sim \sqrt{s} R_V$ soft Higgs bosons are produced via the decay of Higgs field classicalons (Higgsions). The radius $R_V$ of these objects freezes at the Compton wavelength of the Higgs boson $\ell_H = 1/M_H$. Therefore, one probes distances larger than $\ell_w$, instead of a local QFT. In the strongly coupled regime $\sqrt{s} \sim \Lambda_w$, the entire form factor of the WNL Lagrangian should be important. Bound states are expected to form (we call them fuzzyons) that ensure the transition between the Higgs quanta and the Higgsions in the physical spectrum. The little hierarchy $\ell_H \gg \ell_w$ is crucial for the Vainshtein screening to restore unitarity; otherwise, one cannot enter the semiclassical regime of classicalon production that is driven by the leading classicalizing operator.}}
\label{figure_fuzzy}
\end{figure}

Such UV/IR mixing\footnote{This should not be confused with the infamous UV/IR mixing in noncommutative QFT (due to the existence of nonplanar diagrams) that spoils the consistence of gauge theories \cite{Hersent:2022gry}.}, dubbed classicalization \cite{Dvali:2010bf, Dvali:2010ue, Dvali:2010jz, Dvali:2010ns, Dvali:2011nj, Dvali:2011th, Grojean:2011bq, Dvali:2012zc, Dvali:2012mx, Dvali:2014ila, Keltner:2015xda, Addazi:2015ppa, Addazi:2018ivg, Dvali:2020wqi, Addazi:2020nkm, Dvali:2022vzz, Buoninfante:2023dyd} is a generalization of the well-known Vainshtein screening in massive gravity \cite{Vainshtein:1972sx} and some scalar-tensor theories \cite{Babichev:2013usa}. In Ref.~\cite{Dvali:2010ns}, operators like
\begin{equation}
\left( \Phi^\dagger \cdot \Phi \right) \cdot \left( \partial_\mu \Phi^\dagger \cdot \partial^\mu \Phi \right)
\end{equation}
were studied as a source of classicalization in the Higgs sector (cf. Fig.~\ref{figure_fuzzy}), in order to solve the electroweak hierarchy problem, with the production of TeV-scale classicalons of the Higgs field, aka Higgsions, whose phenomenology were studied in Ref.~\cite{Grojean:2011bq}. The remarkable feature is the necessity of a little hierarchy $\Lambda_w \gg M_H$, otherwise the Vainshtein screening cannot develop. Given the lack of evidence for new physics at the LHC to stabilize the EW scale, this is an attractive feature of this scenario. The reader can refer to the original article \cite{Dvali:2010jz} for all the details. Of course, we have just discussed the Higgs self-interactions in this article, and it is important to study if its other interactions also lead to classicalization.

Classicalization is already known to be a nonlocal phenomenon \cite{Keltner:2015xda, Buoninfante:2023dyd}, and its link with other WNL theories has already been suggested in Refs.~\cite{Addazi:2015ppa, Addazi:2020nkm}. The ``transition regime'' $\sqrt{s} \sim \Lambda_w$ is poorly understood because it is deeply quantum and nonperturbative. Instead of the deep-UV regime $\sqrt{s} \gg \Lambda_w$, which is completely determined by the leading classicalizing operator with 2 derivatives \cite{Dvali:2010jz}, the whole tower of higher-derivative operators (from the form factor expansion \eqref{form_factor_exp}) is expected to be important to describe the UV/IR mixing transition. It could be possible that WNL plays an important role in this transition regime, where the specific choice of the form factors could be crucial.

\section{Conclusion \& Outlook}
\label{Conclusion}
In this article, we propose a minimal version of the FSM: a WNL extension of the SM based on the covariant star-product formalism introduced in Ref.~\cite{Nortier:2023dkq}. We generalize the previous article by including non-Abelian gauge symmetries and fermions. This realizes EWSB without introducing ghosts in the physical EW vacuum, which is the main drawback of the previous string-inspired attempts \cite{Gama:2018cda, Hashi:2018kag, Koshelev:2020fok, Nortier:2023dkq}. This minimal FSM has the same approximate global symmetries as the SM. By studying only the Higgs self-interactions, we provide evidence that WNL triggers classicalization in the deep-UV, which deserves to be more deeply investigated in future works.

\section*{Acknowledgments}
Authors thank Hermès Bélusca-Maïto, Luc Darmé, Aldo Deandrea, Gia Dvali, Anish Ghoshal and Nils Marion for useful discussions.

\bibliographystyle{JHEP}

\providecommand{\href}[2]{#2}\begingroup\raggedright\endgroup

\end{document}